\documentclass{egpubl}
\usepackage{eurovis2022}
\usepackage{color}
\usepackage{tikz}
\usepackage{soul}
\usepackage{amsmath}
\usepackage{booktabs}
\usepackage{tabu}

\newcommand{\Cross}{\mathbin{\tikz [x=1.4ex,y=1.4ex,line width=.2ex] \draw (0,0) -- (1,1) (0,1) -- (1,0);}}

\newcommand{\name}[1]{Infographics Wizard}

\SpecialIssuePaper         

\usepackage[T1]{fontenc}
\usepackage{dfadobe}  

\usepackage{cite}  
\usepackage{amssymb}
\BibtexOrBiblatex
\electronicVersion
\PrintedOrElectronic

\usepackage{egweblnk}
\newcommand{\textfloatsp}[1]{
    \setlength{\floatsep}{#1}
    \setlength{\textfloatsep}{#1}
    \setlength{\intextsep}{#1}
    \setlength{\dbltextfloatsep}{#1}
    \setlength{\dblfloatsep}{#1}
}

\textfloatsp{+2mm}

\title{\name{}: Flexible Infographics Authoring \\and Design Exploration}

\author[Tyagi \textit{et al.}]
{\parbox{\textwidth}{\centering Anjul Tyagi$^{1}$\orcid{0000-0003-3822-124X}
        Jian Zhao$^{2}$\orcid{0000-0001-5008-4319}
        Pushkar Patel$^{3}$\orcid{0000-0003-3410-8856}
        Swasti Khurana$^{3}$\orcid{0000-0002-8470-3082}
        Klaus Mueller$^{1}$\orcid{0000-0002-0996-8590}
        }
        \\
{\parbox{\textwidth}{\centering $^1$Computer Science Department, Stony Brook University, New York, USA\\
        $^2$Cheriton School of Computer Science, University of Waterloo, Ontario, Canada \\
        $^3$Indian Institute of Information Technology Vadodara, Gujarat, India \\ 
       }
}
}

\begin{document}
\maketitle
\begin{abstract}
   Infographics are an aesthetic visual representation of information following specific design principles of human perception. Designing infographics can be a tedious process for non-experts and time-consuming, even for professional designers. With the help of designers, we propose a semi-automated infographic framework for general structured and flow-based infographic design generation. For novice designers, our framework automatically creates and ranks infographic designs for a user-provided text with no requirement for design input. However, expert designers can still provide custom design inputs to customize the infographics. We will also contribute an individual visual group (VG) designs dataset (in SVG), along with a 1k complete infographic image dataset with segmented VGs in this work. Evaluation results confirm that by using our framework, designers from all expertise levels can generate generic infographic designs faster than existing methods while maintaining the same quality as hand-designed infographics templates.


\end{abstract}  
\section{Introduction}\label{s:introduction}
\begin{figure*}[tb]
    \centering
        \includegraphics[width=\textwidth]{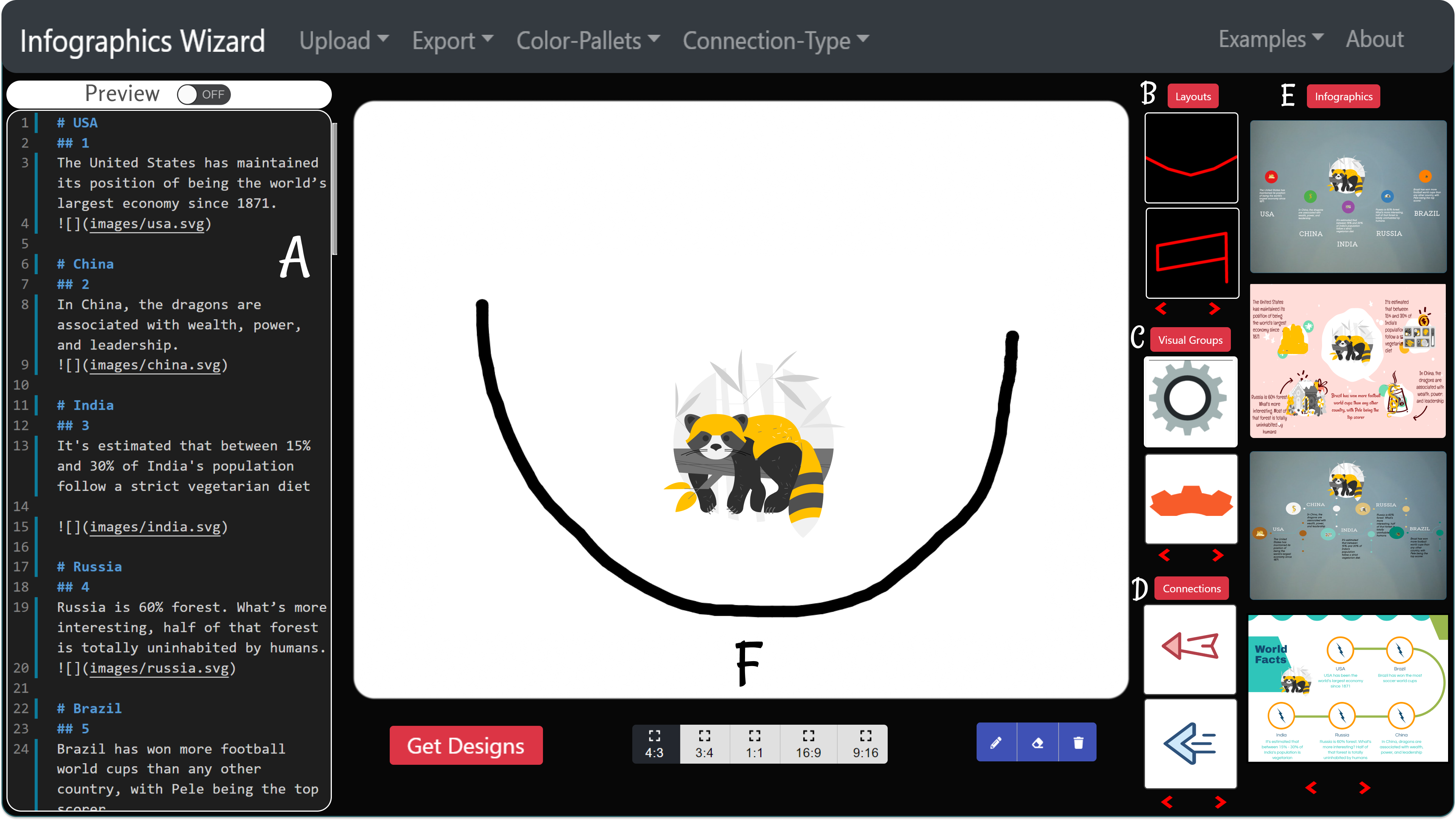}
        \vspace{-6mm}
    \caption{\name{} implements a flexible framework for full- and semi-automated infographic generation. Based on user’s input in the markdown format (A), \name{} generates various infographic design recommendations (F). Users can further customize the infographics by altering different design components via the recommendation panel for the Visual Information Flow layouts (B), the design of individual Visual Groups (C), and the connecting elements between Visual Groups (D).
More experienced designers can optionally provide a main pivot graphic, the general information flow, and custom designs on a canvas (F) via direct manipulation to control the generation and recommendation of infographic design.} 
    \label{fig:teaser}
\end{figure*}
Infographics have been widely used in various areas like fashion, advertisement, and business to convey complex data-driven narratives aesthetically by following well-studied design principles of human perception \cite{lankow2012infographics, dick2014interactive}. 
Infographics cover a vast range of designs, however, the most common types have some common patterns as pointed in the past research. To convey information about a particular idea, designers generally divide infographics into separate repeated design components---\textit{Visual Groups (VGs)}---each carrying a specific piece of information \cite{lu2020exploring}. 
These design pieces are then organized together in a logical sequence---\textit{Visual Information Flow (VIF)}---which forms an infographic \cite{lu2020exploring}. However, it is challenging for non-experts to develop compelling infographic designs because of the required design experience. 
To address these challenges, several fully-automated infographic design tools like Text2Viz \cite{cui2019text} and DataShot \cite{wang2019datashot} exist. 
While convenient, these tools are often based on a limited set of templates, lacking the variety of designs they can generate.  
Semi-automated tools allow the user to apply their design ideas with some cues and suggestions (e.g., \cite{brehmer2019timeline,chen2019towards,wang2018infonice,coelho2020infomages}). However, most of the proposed techniques either focus on timeline infographics or charts. 

\begin{figure*}[tb]
    \centering
        \includegraphics[width=\textwidth]{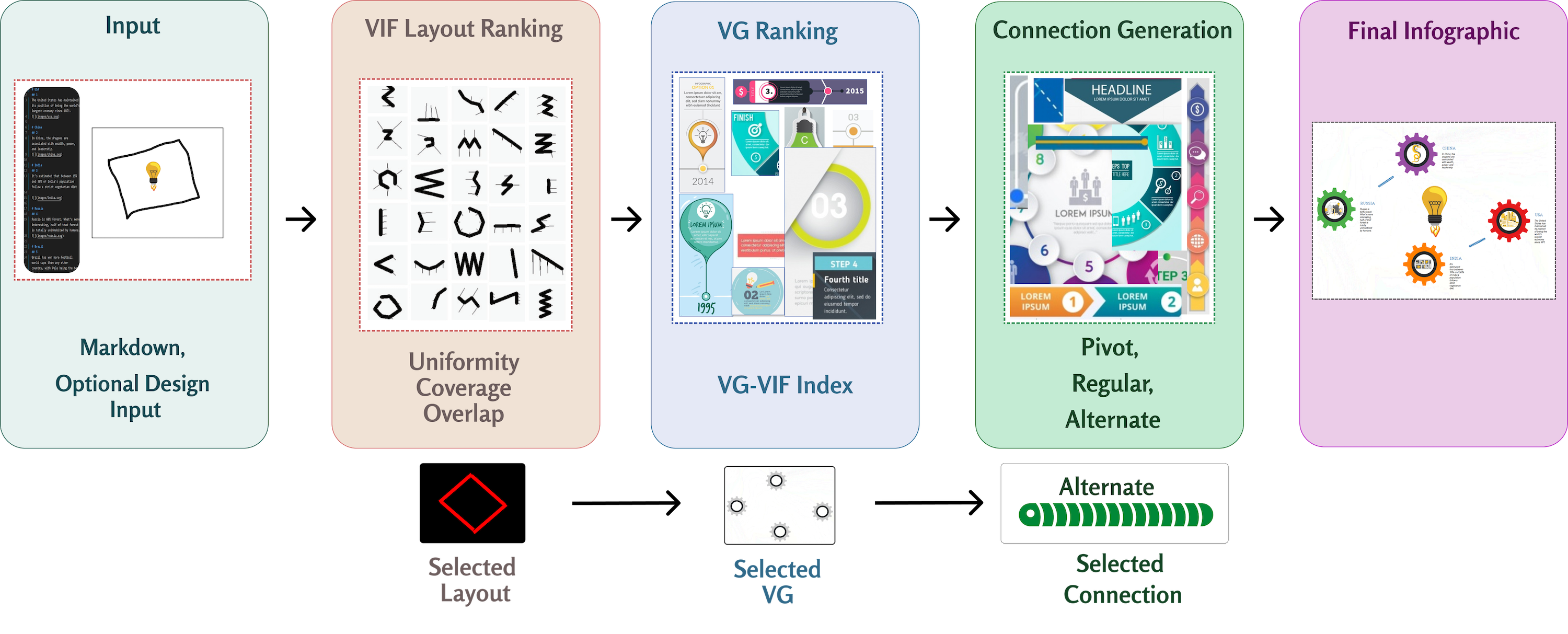}
        \vspace{-9mm}
    \caption{The three-stage pipeline of our framework for infographic generation (See Section \ref{s:infographic_generator} for details). The first stage recommends the VIF layouts, 
    The second stage selects the VG designs, and the third stage generates connections. This example shows a very simple infographic skeleton.  Users can customize and add semantic components to the infographic by modifying background, pivot graphic, colors, and adding custom VIF, VG, and Connection designs. All these functions are user-controlled while the basic skeleton generation (VIF, VG designs, and Connection style and designs) are automated with our framework.}
    \label{fig:pipeline}
\end{figure*}

In this work, we use a holistic view of infographic design by leveraging the concepts of VIF and VG \cite{lu2020exploring}.
We propose a general framework derived from a formative study with 10 expert designers. This decouples the overall structure of infographics and offers support for flexible design automation while providing the freedom to control major information pieces. 
Specifically, with the help of designers with different expertise levels, we discovered four main design components in infographics, including (1) \textit{VIF layouts} that represent the backbone structure of the infographic, (2) \textit{VG designs} that determine repeating design components holding a specific piece of information, (3) \textit{pivot graphics} that set the stage of other design components and the overall infographic, and (4) \textit{connecting elements} that bind individual VGs together (or to the pivot graphic). We combined these main design components to create a generic infographic design pipeline discussed in Figure~\ref{fig:pipeline}.
To the best of our knowledge, this work is the first general solution towards automating the infographic design process.

Infographic design elements can be classified into two categories: (1) generic and (2) domain-specific (e.g., semantics related), which reflects different design needs. Decoupling infographic design with our framework allows each part to be individually optimized in fully- and semi-automated authoring tools. Our framework automates the design of generic components of the infographic (VIFs, VGs, and Connections), while also supporting customization of the domain-specific components (e.g. images, pivot elements, custom VG, connection designs, background, and colors) which are fully controlled by the users. 
To foster future research with the framework, we will also contribute an infographic dataset of 1K images with annotated and extracted VG designs collected using Amazon Mechanical Turk \cite{mturk}. 

We operationalize this framework by developing an interactive tool, called \textit{\name{}}, for rapid prototyping and design exploration of infographics (Figure~\ref{fig:teaser}). 
Inspired by the prevalence of markdown languages as seen in web development and Jupyter Notebooks, \name{} allows the user to separate the manipulation of content and presentation in infographics design.   
In particular, the user specifies the content of each VG of the designed infographic in a markdown format (Figure~\ref{fig:teaser}A), without worrying about the layout or appearance.
Then, \name{} fulfills the framework design by recommending appropriate infographic designs (Figure~\ref{fig:teaser}E) along with the VIF layouts (Figure~\ref{fig:teaser}B), VG designs (Figure~\ref{fig:teaser}C), and connecting elements (Figure~\ref{fig:teaser}D). 
We validated \name{} through a four-stage multi-aspect evaluation with the results indicating the effectiveness of our framework and \name{} in improving the design process of infographics. In summary, our contributions in this paper include\footnote{The code, datasets, and supplementary materials of this work can be accessed via \url{https://tyagi-iiitv.github.io/blog/2022/04/infographics-wizard}.}:
\begin{itemize}
    \item A general, extensible framework for infographic design that captures standard design components and workflows (Figure~\ref{fig:pipeline});
    \item An interactive tool, \name{}, that implements the framework to provide automatic and semi-automatic generation of infographics based on flexible user inputs and manipulation;
    \item A dataset of 1k infographic images with labeled Visual Groups, and extracted individual Visual Group design SVGs and connection SVGs;
    \item Results and analyses from a comprehensive evaluation of our approach on various aspects.
    
\end{itemize}

 
\section{Related Work}
\label{s:related_work}
 
\subsection{Infographics and Relevant Studies}
\label{ss:infographics}
The design of infographics depends on several factors related to human perception of information, which makes them more memorable and engaging \cite{bateman2010useful,haroz2015isotype,harrison2015infographic}. Some previous work has focused on understanding the use of visuals in infographics and its role in making the infographics more memorable and visually pleasing \cite{borkin2013makes,harrison2015infographic,skau2017readability, bylinskii2017understanding, madan2018synthetically, Fosco2020, byrne2015acquired, harrison2015infographic, lan2021smile}. For example, Bateman \textit{et al.} \cite{bateman2010useful} discussed infographic design in terms of increased memorability. 
Chen \textit{et al.} \cite{chen2019towards} analyzed timeline infographics and proposed some design component concepts specific to timelines. Zheng \textit{et al.} \cite{zheng2019content} proposed a fully automated deep learning-based approach to generate magazine layouts.
More generally, Lu \textit{et al.} \cite{lu2020exploring} summarized infographic designs based on the notions of Visual Information Flow (VIF) and Visual Group (VG). They clustered the infographic design space into 12 categories based on VIF; however, they did not fully investigate the design of VGs. 


Our contribution lies in automating the infographic design workflow and extending the infographic design concepts (Section \ref{s:design_study}). Although these studies help evaluate, understand, and categorize existing infographic designs, the goal of automated infographic generation is out of scope in all of these works, except in Chen \textit{et al.}'s that focuses only on timeline infographics. Also, compared to deep learning techniques of automating infographic design \cite{zheng2019content}, our approach is more straightforward, faster, and explainable. 

\subsection{Infographic Generation Tools}
\label{ss:ig_tools}
Infographic generation tools can be classified into three main categories: manual, semi-automated, and fully automated. 
The manual techniques include design tools that provide complete control to designers for authoring every aspect of infographics from scratch, such as Adobe Illustrate and other commercial software \cite{photoshop, marvel, proto, mockflow, adobexd}. On the other hand, the semi-automated design tools allow for easier infographic generation while keeping some partial control to designers in the process. For example, tools like Proto.io \cite{piktochart}, and Timeline Storyteller \cite{brehmer2019timeline} support the generation of timeline infographics given custom time-series data from existing templates. Chen \textit{et al.} \cite{chen2019towards} improved the structure of timeline infographics by automatically extracting templates from infographic images using deep learning. 
Finally, fully automated tools directly generate infographics from input data or resources, such as the ``Design Idea'' function of Microsoft PowerPoint. InfoColorizer recommends color palettes for infographics via a data-driven approach \cite{Yuan2021}. Also, DataShot automatically generates fact sheets based on existing templates by automatically extracting data facts from a custom dataset \cite{wang2019datashot}. Another work by Cui \textit{et al.} \cite{cui2019text} produces infographics automatically from text using pre-existing templates for simple proportion-related descriptions. 

Although all of these works provide different levels of control to the designers, there are limitations to each category. The manual tools have a steep learning curve. 
The semi-automated approaches focus on automating low-level designs 
but they only support timeline infographics or charts. 
Fully automated techniques have the fastest turnaround time for generating infographics, but they are limited by prescribed whole-infographic templates. 
We use a holistic view of infographic designs and propose an extensible framework, bridging the gap between semi-automated and fully automated techniques. \name{} automates the general designs of infographics (VGs, VIFs, and Connections) while users can control the semantic designs of the infographics (background, pivot elements, custom VG and VIF designs, and color schemes). 

\subsection{Visualization Recommendation}
\label{ss:vis_reco}
Besides the work related to infographic generation, there is considerable research in developing recommendation systems for data visualizations (charts). These techniques can be broadly classified into two categories: rule-based~\cite{steele2011designing, mackinlay1986automating,roth1994interactive, tyagi2019ice,cao_graphs} and data-driven techniques~\cite{hu2019vizml,dibia2019data2vis,dibia2019data2vis,satyanarayan2016vega,kumar2019task,tyaginas}. The rule-based techniques \cite{steele2011designing, mackinlay1986automating} introduced methods to generate a latent space of charts using compositional algebra, which was later improved by SAGE \cite{roth1994interactive}. To support efficient search in this latent space of charts, CompassQL \cite{wongsuphasawat2016towards}, later improved by Voyager \cite{wongsuphasawat2015voyager} and Voyager 2 \cite{wongsuphasawat2017voyager} were developed using query specifications on this search space. In the data-driven techniques, VizML \cite{hu2019vizml} uses machine learning to develop encodings representing the relationship between data characteristics and visualizations. Data2Vis \cite{dibia2019data2vis} is another deep learning framework to directly generate visualizations given the data using the visualization grammar introduced in Vega \cite{satyanarayan2015reactive} and VegaLite \cite{satyanarayan2016vega}. 

While these systems are suitable for ranking data visualizations, they only work for charts. Infographics have significantly different characteristics than standard charts, and generating and ranking them is inherently different from working with charts, which is the focus of our work.
\section{Infographics Generation Framework}
\label{s:framework}

In this section, we introduce our flexible framework for infographics generation, which is distilled from a formative study with participants with different levels of experience in infographic design. Example infographics generated using the tool (see Section \ref{s:infographic_generator}) based on our framework are shown in the supplementary material. 


\subsection{Formative Study}
\label{ss:survey}

To systematically evolve our idea of an infographic generation framework, we first conducted a formative study to get to know user requirements, their views of infographic designs, and general workflows. 
This approach helped concertize our framework and tool design with a user-centered evaluation at an earlier development stage. 
The formative study participants were carefully chosen to be designers and researchers working in data visualization and infographic design, with various experience levels. Out of ten participants, two were professional designers working in the industry, two were professors working in data visualization, three were Ph.D. students working in data visualization, and three were undergraduate students in Computer Science interested in data visualization.

The participants were initially introduced to infographics and the target we try to achieve. We asked how they would conceptually model infographic designs. Next, they were introduced to the VIF and VG concepts, followed by discussions about existing infographic generation tools and their shortcomings. 
We further inquired how they would use the existing tools to design infographics and the ideal tools they could imagine. 
This helped us devise a new framework that could flexibly support automated infographic generation and recommendations for various cases.

\subsection{Key Findings - Infographics Design Components}
\begin{figure}[tb]
    \centering
        \includegraphics[width=\columnwidth]{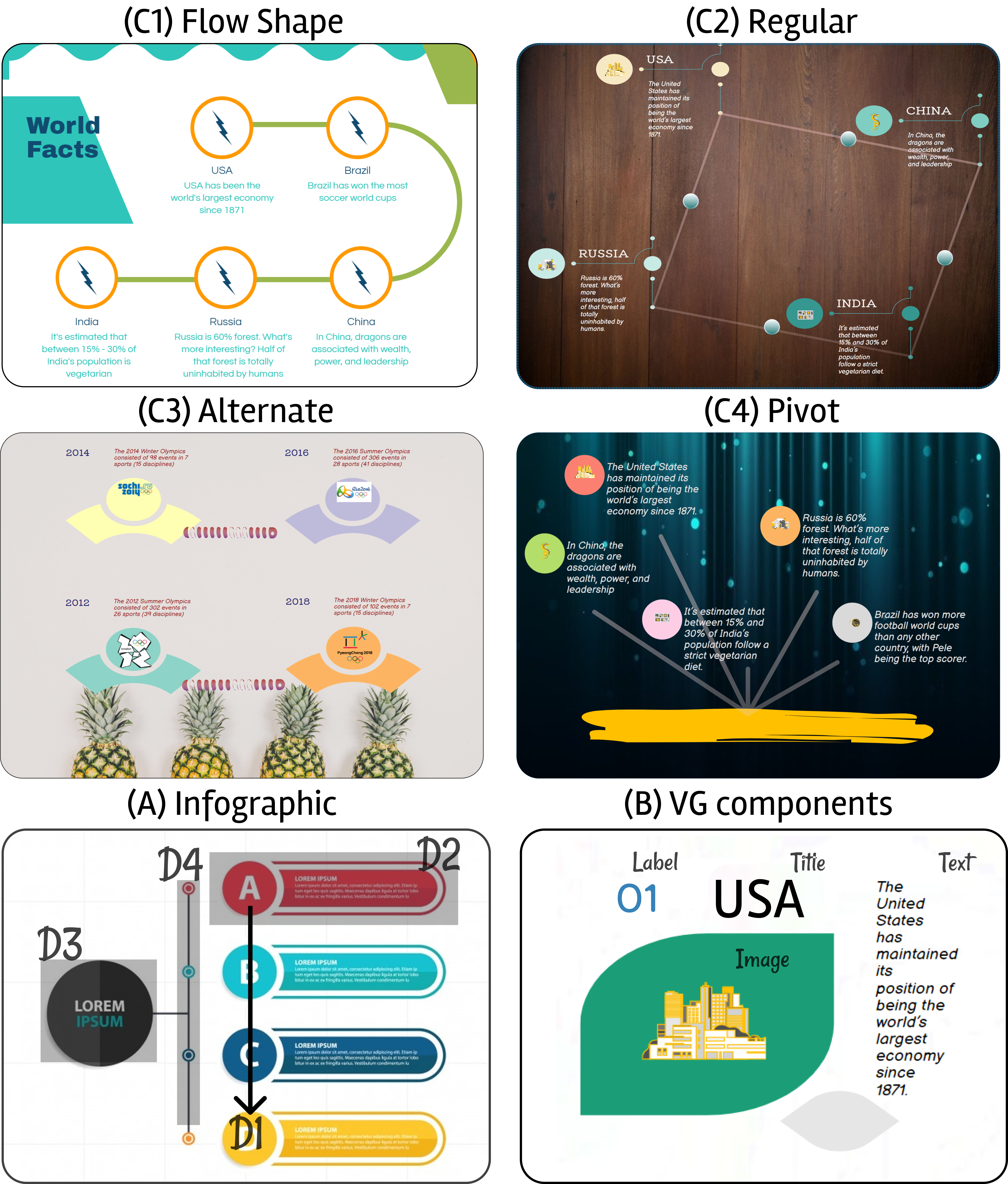}
        \vspace{-6mm}
    \caption{\textbf{C1-C4:} Four types of connection styles supported in our framework, see Section \ref{ss:components} for details. \textbf{(A)} Design components of an infographic proposed in the formative study (Section \ref{ss:survey}).
    \textbf{(B)} Individual components inside a VG. }
    \label{fig:connections_and_vgs}
\end{figure}

\label{ss:components}


To find the most common design patterns in infographics, participants used a process similar to Text-to-Viz \cite{cui2019text}. They collected the top 200 unique infographics using Google Image search with the keyword ``infographic" and consolidated a list of top design patterns. Each of these design patterns (discussed below) were then verified to be occurring in at least 82\% of the search results, thus validating the coverage requirements for developing a generic infographic design framework (see Figure \ref{fig:connections_and_vgs}-A). The design patterns are as follows:

\textbf{D1: Visual Information Flow Layouts.} 
The participants confirmed that the concept of VIF reflects the backbones of the story or information that an infographic aims to express, thus validating the findings from Lu \textit{et al.} \cite{lu2020exploring}. Participants further agreed that allowing designers to optionally control the VIF layout intuitively is an essential component of any infographic generation workflows.

\textbf{D2: Visual Groups Designs.} 
The participants also echoed the concept of VG, and they stated that VGs could be designed separately and reused in one or multiple infographics. These VGs could then be placed according to the VIF to generate an infographic. 
They also mentioned that it is essential to mitigate the effort in VG design for novices while allowing them to fully control the content of VGs and the overall structure of an infographic. The participants further mentioned that separating the content and the design, like CSS and HTML for web design, could be of great benefit.  

\textbf{D3: Pivot Graphics.} 
The participants pointed out that some infographics contain a pivot graphic, which are design elements different from VGs, acting as a central background component or binding anchor for an infographic. This coincided with the ``main body'' of timeline infographics in Chen \textit{et al.}'s analysis \cite{chen2019towards}. Participants observed that such pivot graphics exist in a wide range of infographics in addition to timelines and stated that supporting the customization of this element is needed in infographics design when appropriate.

\textbf{D4: Connecting Elements.} 
The participants formulated the connections between VGs, sometimes between a pivot graphic and different VGs, as one of the critical infographics components. The connecting elements cannot be included in VGs because they are decorations not contributing to the content semantics of the VGs, and sometimes there is no one-to-one mapping between the connections and the VGs. 
To support infographic generation, we categorized the connecting elements into four classes: flow-shaped, regular, alternating, and pivot connections, based on our discussion with the participants. These types are discussed below with an example shown in Figure \ref{fig:connections_and_vgs}.

Flow Shape Connections: These connections are placed around the center of the infographic or around the pivot element (if present). Following the direction of the infographic's VIF layout, their placement is decided based on the placement of VGs on the VIF flow, and the slope is decided by the corresponding flow line connecting two VGs.

Regular Connections: These connections follow the direction of the VIF layout of the infographic and are placed at the center of each flow line. The placement angle follows the slope of the flow line and the length of each connection depends on the distance between corresponding VGs in the infographic. 

Alternate Connections: These connections are similar to regular connections but are placed at alternating flow lines. Like regular connections, the angle and position of these connections depend on the flow line's center and the slope, and the length depends on the distance between VGs. 

Pivot Connections: In case the infographic has some pivot graphics, these connections are generated from the center of the pivot element towards all the VGs. For a particular VG, the connection placement is in the center of the line joining the pivot element and the VG. Also, the connection angle is the slope of connecting line, and the length is decided based on the distance of the VG from the pivot element. 


Each of the components is critical for designers to control for generating customized infographics. Various levels of controls result in a flexible workflow with fully- or semi-automated infographic development.

\subsection{Methodology}
\label{s:overview}
Based on the design components discussed above, we construct a pipeline-based framework with three main stages, as shown in Figure~\ref{fig:pipeline}. Input to the first stage is the user-provided information, which is used to decide the number of VGs required to create an infographic. Shown as \textit{Input} in Figure~\ref{fig:pipeline}, the information can be provided in the form of markdown text, where each bullet point can include a title, text, label, and an image. The user can optionally provide the pivot graphics like images and layout drawings using a canvas-based interface. This information is then passed onto the following stages (refer to Section \ref{s:infographic_generator} for details of each stage):

\textbf{S1: Fitting Visual Information Flow Layouts.} 
This is the first stage where the existing VIF layouts extracted from infographic datasets are ranked based on the pivot graphics (optional), canvas size, and the user's VIF sketch input (optional). We discuss the details of our curated dataset in Section \ref{s:dataset}.

\textbf{S2: Selecting Visual Group Designs.} 
Once the VIF layouts are ranked, the next step is to select the VG designs that best fit the ranked VIF layouts. We curated a VG-VIF index for this purpose, explained in Section \ref{s:infographic_generator}, providing final VG designs that can generate the infographic. 
The user can alter the proposed designs by choosing from multiple recommendations or going back to the previous stage to provide new inputs. Once the VG design is finalized, it is placed based on the VIF layout, rotated, and scaled to fit the infographic's size and design.

\textbf{S3: Generating Connections.} After finalizing the VIF layouts and VG designs, the final step is to connect the pivot elements (if any) and the VGs. Similar to the VG-VIF index, we also created a C-VIF index to assist in ranking connections based on the VIF layouts (details in Section \ref{s:infographic_generator}). 

This three-step process generates the basic skeleton of an infographic based on the rules discovered in the formative study. Users can further customize several aspects of the final design by altering the background, pivot graphic, colors, and custom VG, VIF, and connection designs. 
Our framework is a step towards bridging the gap between the existing fully- and semi-automated infographic generation techniques. 
Novice designers can take advantage of the fully automated design pipeline to explore a vast set of infographic design recommendations, while more experienced designers can use our framework to explore a set of designs possible based on some design constraints. Designers can also export the generated infographics as SVG files to fine-tune and alter very low-level design elements.  
\section{Infographics Dataset with Visual Groups}
\label{s:dataset}


As mentioned previously, our framework, or any automated or semi-automated infographics generation tools, need to be driven by existing infographic designs created by experts, thus adequately leveraging the collective wisdom, aesthetics principles, and perceptual rules.
However, there is still a lack of high-quality infographics datasets with fine-grained and accurate annotations based on the design principles proposed in Section \ref{ss:survey}.

One of our main contributions in this work is the infographics dataset of 1K images with annotated VG designs. The main goal to curate this dataset was to compensate for the lack of hand-designed VG datasets. Since VGs in infographics are a reasonably new idea, recently introduced by Lu \textit{et al.} \cite{lu2020exploring}, there does not exist any dataset targeting specifically towards annotating the VGs in infographics. Creating this dataset helped us extract VG designs to extend the scope of infographics that can be generated with our framework. Along with the 1K human segmented VGs in complete infographic images, we also release the extracted VGs from each of these images in separate SVG files. These designs are used in our framework to generate infographics. We annotate the VGs with both a segmentation mask and a bounding box.

\textbf{Source Images.}
We used the results by Lu \textit{et al.} \cite{lu2020exploring} to sample 1000 infographics belonging to each of the 12 VIF categories (as proposed in their work \cite{lu2020exploring}) and in the ratio as they appear in the original dataset (Figure 8 in \cite{lu2020exploring}). Because of our framework's design aspect, since we aimed to relate the VG designs with VIF layouts, sampling infographic images based on the VIF layout was crucial for consistency with real-world designs.  

\textbf{Processing the Segmentation Maps.}
Human-generated segmentation maps are coarse with ill-defined shapes. To generate a well-defined VG segmentation map, we employed the automated GrabCut \cite{rother2004grabcut} algorithm, similar to \cite{chen2019towards}. The human annotations were used as input to the algorithm to generate high-quality segmentation masks for VGs. This mask was then passed as input to Solaris \cite{solaris} which generated SVG paths for segmented VG designs. Finally, we added additional components to the SVG based on bounding box annotations of individual VGs, provided with the dataset by Lu \textit{et al.} \cite{lu2020exploring}.

\section{Design Components Extraction} 
\label{s:design_study}

Based on the formative study, to create a working prototype of the proposed framework, we require individual SVG design components of VGs and Connections, along with extracted VIF layouts information (\textbf{D1, D2, D4}). Using the annotations from our collected dataset (discussed in Section \ref{s:dataset}), we extracted the VIF, VG, and connection designs from the existing infographics datasets. 



To extract the VIF from our infographics dataset, we used the VIF extraction algorithm discussed in~\cite{lu2020exploring}.
For VG extraction, we chose 
three sources; first is Adobe Stock \cite{adobe}, which contains SVG designs of several human-generated infographics, from which we manually separated the VG designs; the second source is our human segmented VG dataset which we curated for this work, explained in Appendix A; the third source is taken from the work by Chen \textit{et al.} \cite{chen2019towards} where we extracted VGs from timeline infographics using a Mask R-CNN \cite{he2017mask}. 
For connection dataset, we collected designs from adobe stock similar to the collected 200 infographics from the formative study. All connections from these images were found on the portal, which mainly were simple shapes (like arrow, line, circle, etc.) which could be modified based on certain conditions. Details about how each of these components were extracted are discussed in Appendix B. 
\section{Infographics Wizard}
\label{s:infographic_generator}

Following the indicated pipeline of our framework (see Section \ref{s:overview}), we developed an interactive tool, \name{}, that provides infographic recommendations of different design components in parallel for rapid prototyping and exploration of infographics. An overview of the tool is shown in Figure \ref{fig:teaser}. 
This section discusses the algorithms we employed to fulfill each pipeline stage of our proposed framework. Note that \name{} is a specific realization of our framework, where other suitable algorithms and datasets can be employed to fulfill the proposed pipeline. 


\textbf{Visual Information Flow Layout Recommendation. }
\begin{figure}[tb]
    \centering
    \includegraphics[width=\columnwidth]{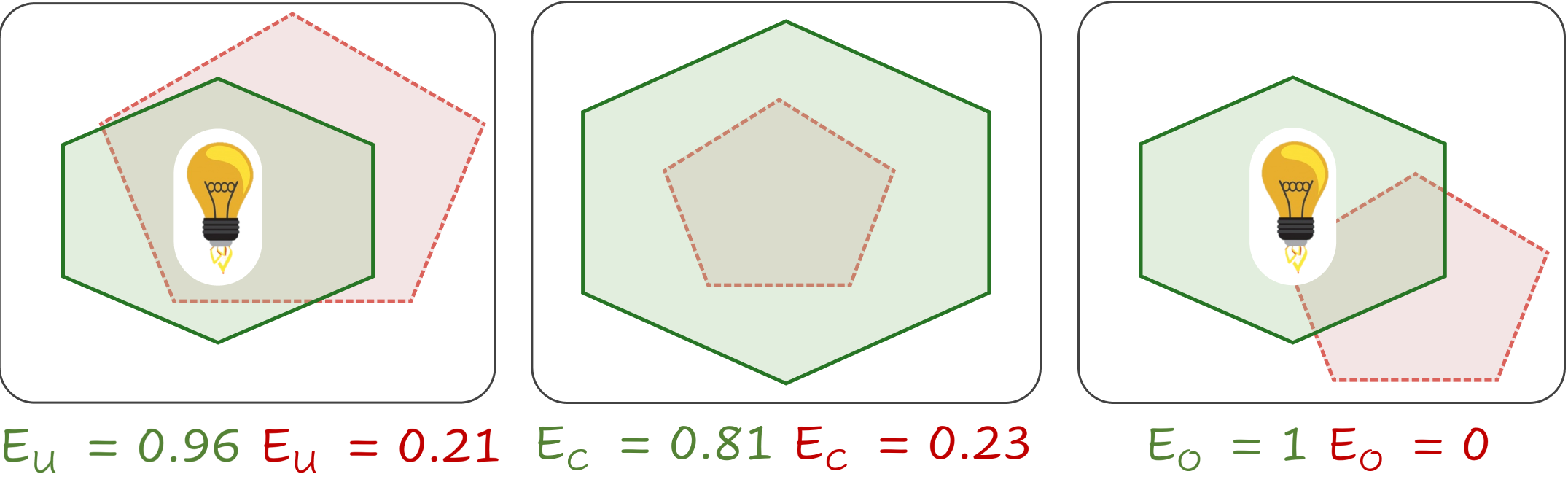}
    \vspace{-6mm}
    \caption{The VIF layouts ranking energy function demonstrations.}
    \label{fig:layout}
\end{figure}
This algorithm finds the best layouts from our extracted VIF layouts dataset, given the user design constraints. We did not optimize for layout generation directly and instead chose a ranking-based approach because retrieval of related existing VIF layouts is faster and less resource-intensive than directly optimizing for finding the best layout \cite{o2015designscape, o2014learning}. We tested our approach for upto 12 inputs using our dataset. With the ranking-based approach, we can recommend relevant layouts in real-time and instantly follow user feedback. 
The ranking of VIF layouts are scored based on the energy function shown in Equation~\eqref{eq:layout_rank} by $E_{L}$
\begin{equation}
    \label{eq:layout_rank}
    E_{L} = E_{O}(\alpha E_{C} + (1-\alpha) E_{U})
\end{equation}
where $\alpha \in [0,1]$, balances the contribution of $E_{U}$ and $E_{C}$ in the final energy function. $E_{U}$, $E_{O}$ and $E_{C}$ are the three components of the energy function. $E_{U}$ gives the score to uniformity, meaning how uniformly the VIF vertices are located from the pivot element center. $E_{C}$ scores the coverage based on how much the convex hull of a VIF covers the canvas area. $E_{O}$ is the overlap score, which is $0$ if there is an overlap between any VIF vertex and the pivot element, and $1$ otherwise. 

In some cases, designers might provide an initial sketch of a coarse layout, for example, shown in Figure \ref{fig:teaser} and \ref{fig:pipeline}. In such cases, since we already have a VIF layout, we aim at finding the nearest neighbors to this user-provided VIF layout from our dataset. To calculate the nearest neighbors, we first calculate the extreme points on the hand-drawn contour using the technique by Teh \textit{et al.} \cite{teh1989detection}, which has been extensively used in many applications. These points give us the estimated positions of VGs, which are then matched with the existing database of VIF layouts to find the closes neighbors of these extreme points. 

\textbf{Visual Group Design Recommendation. }
Based on our infographic generation pipeline, after ranking the VIF layouts, the next step is to choose corresponding VG designs that best fit the selected VIF layout. To support this, we develop a \textit{VG-VIF index} that provides a method to rank VGs given a VIF layout from our dataset. The VG-VIF index aims to capture a global relationship between the VG designs and the VIF layouts to support a ranking system that allows accurate recall of the suitable VG designs given a VIF layout. The VG scores are obtained from the VG-VIF index, which shows how well-fitting a VG is for a given layout. These scores are sorted, and a subset of high-scoring VGs are selected, which match the user's markdown input. Some fine-tuning is applied - like rotation and scaling to properly place the VGs on the infographic. 

\textbf{Connection Recommendation and Filtering. }
The connections in an infographic refer to the design components used to connect VGs and pivot graphics with respect to the VIF layout (\textbf{D3, D4}). For ranking of connection styles, we categorize the connection ``styles'' into five classes, of which four were discussed in Section \ref{ss:components}, and 1 class ``none'' was added, which means that the infographic has no connecting elements. The goal is to establish the relationship between the infographic theme and connect elements by ranking these five classes of connection styles based on the VIF layout. For this purpose, we manually created a ``C-VIF'' index to store the connection styles and corresponding VIF layouts from the 200 infographics chosen during the formative study and an extra 200 infographics from the existing template infographics dataset \cite{lu2020exploring}. These 200 infographics from Lu \textit{et al.} \cite{lu2020exploring} were carefully chosen to equally represent all the 12 VIF categories proposed in their work.
For each of these connection designs, we randomly picked the connection designs to be shown to the user on \name{}, shown in Figure \ref{fig:teaser} (D). The user can choose to browse more connection designs or pick a connection design, while the tool automatically controls the best connection styles to use in the recommendations. The designers also have the freedom to control the connection style still if required. Some more details about these methods are provided in Appendix C. 




\section{Evaluation}
\label{s:evaluation}

We evaluated our framework and the interface \name{}, by following the schemes discussed in the nested model for visualization interface design \cite{munzner2009nested}, which includes a comparison study, case studies, an in-lab user study, and a survey study. 



\subsection{Comparison Study}
\label{ss:ach}
Based on our application scenario, we compare the existing infographic generation tools listed in Section \ref{s:related_work} and \name{} in terms of design functionalities. The participants from the formative study proposed a set of design functionalities for comparison as they are a sample of the target audience for \name{}. The hypotheses are as follows (H1-H5):

\begin{table}[tb]

  \caption{Comparing \name{} and existing infographic generation tools based on the design functionalities formulated during the formative study, discussed in Section \ref{ss:survey}. A checkmark means the proposed functionality is satisfied by the tool and the cross mark means the functionality is not satisfied. The result column shows the tools which satisfy all the proposed functionalities.}
  \label{tab:comp_hyp}
  \small
  \begin{tabular}{p{0.35\columnwidth}  p{0.04\columnwidth}  p{0.04\columnwidth}  p{0.04\columnwidth}  p{0.04\columnwidth}  p{0.04\columnwidth} p{0.1\columnwidth} }
  \toprule
  \textbf{Design Tool} & H1 & H2 & H3 & H4 & H5 & \textbf{Result}\\
  \midrule
  Chen et al. \cite{chen2019towards} & \checkmark & \checkmark & $\Cross$ & $\Cross$ & $\Cross$ & \textbf{$\Cross$} \\
  \midrule
  Timeline Storyteller \cite{brehmer2019timeline} & $\Cross$ & \checkmark & $\Cross$ & \checkmark & \checkmark & \textbf{$\Cross$} \\
  \midrule
  Text-to-Viz \cite{cui2019text} & \checkmark & $\Cross$ & $\Cross$ & $\Cross$ & \checkmark & \textbf{$\Cross$} \\
  \midrule
  VIF Flows \cite{lu2020exploring} & $\Cross$ & \checkmark & $\Cross$ & $\Cross$ & \checkmark & \textbf{$\Cross$} \\
  \midrule
  DataShot \cite{wang2019datashot} & \checkmark & \checkmark & $\Cross$ & \checkmark & $\Cross$ & \textbf{$\Cross$} \\
  \midrule
  Adobe Illustrator, MS Powerpoint & \checkmark & $\Cross$ & $\Cross$ & $\Cross$ & \checkmark & \textbf{$\Cross$} \\
  \midrule
  SmartArt Powerpoint & \checkmark & $\Cross$ & $\Cross$ & $\Cross$ & \checkmark & \textbf{$\Cross$} \\
  \midrule
  Design Tools \cite{adobexd, photoshop, marvel, mockflow, piktochart, proto} & $\Cross$ & \checkmark & \checkmark & \checkmark & $\Cross$ & \textbf{$\Cross$} \\
  \midrule
  \textbf{\name{}} & \textbf{\checkmark} & \textbf{\checkmark} & \textbf{\checkmark} & \textbf{\checkmark} & \textbf{\checkmark} & \textbf{\checkmark} \\
 \bottomrule
  \end{tabular}%
\end{table}


\textbf{H1: Separate Design from Content.} The tool allows automatically handling the design component of infographics generation, with designers only having to control the infographics' content. 

\textbf{H2: Overall Layout Design.} The tool supports authoring and exploring multiple layouts, which can be used to design infographics from the given user content.

\textbf{H3: Visual Group Design.} The tool supports manipulating infographic designs on a VG level. Designers should be able to explore various VG designs and also customize infographics based on a user-provided VG design. 

\textbf{H4: Connection Design.} Like H3, the tool allows generating and exploring different connection designs and styles in infographic designs. The designers should optionally be able to remove any connections if required. 

\textbf{H5: Recommendations.} Based on H1, the tool supports ranking different infographic designs given the content and design inputs. These rankings should provide an exploration of relevant infographic designs based on elements extracted from existing datasets.  

To the best of our knowledge, comparing existing tools according to Table \ref{tab:comp_hyp} shows that our tool satisfies all of the proposed infographic design functionalities. Using our framework, \name{} supports flexible infographic designing with user input that combines the benefits of fully- and semi-automated features.

\subsection{Demonstrating Design Capabilities with Example Cases}
\begin{figure*}[tb]
    \centering
        \includegraphics[width=\textwidth]{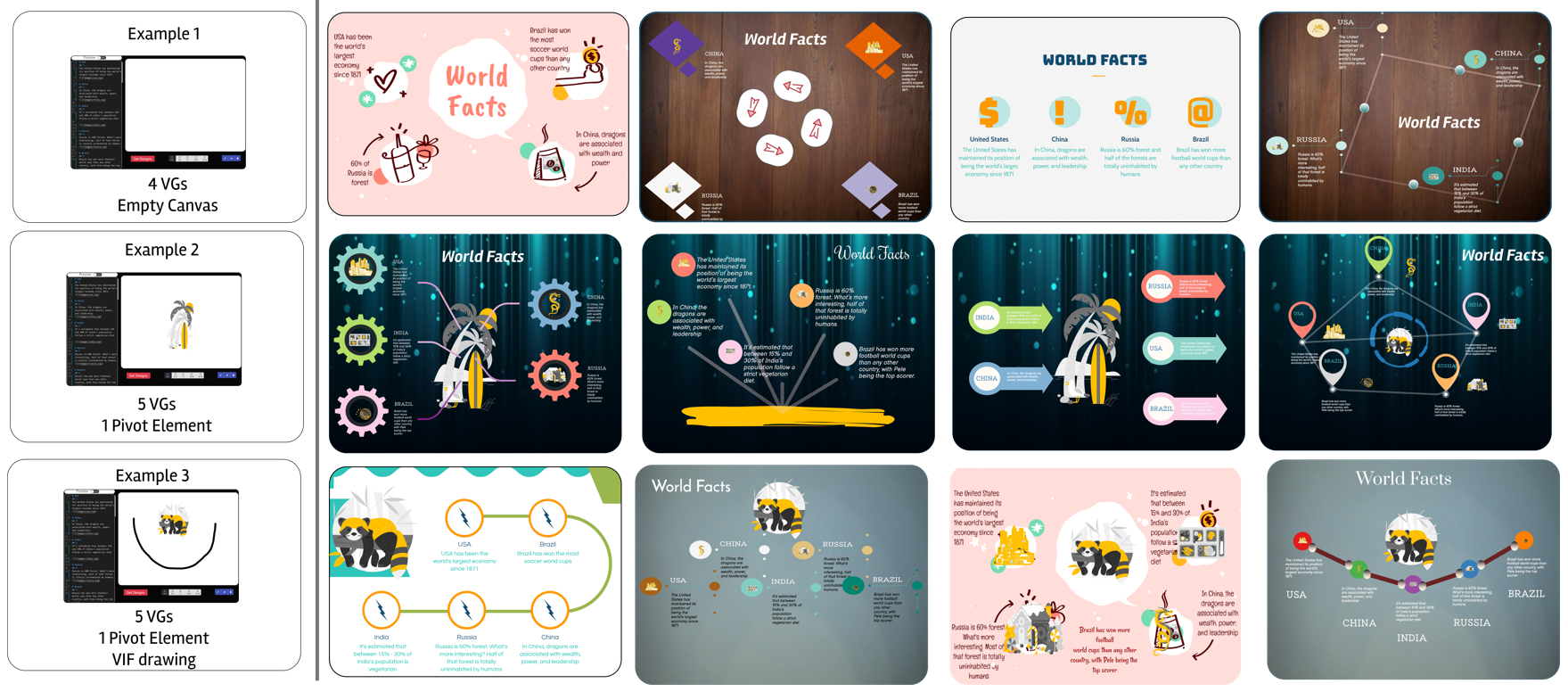}
        \vspace{-8mm}
    \caption{Example infographics generated using our framework using the example cases. We evaluated our framework with three example inputs, each of which had different content and design, shown on the left of each row. For each input, our framework recommended infographics, where we show some recommended infographics in each row for similar inputs.} 
    \label{fig:example_cases}
\end{figure*}

We demonstrate our framework's design capabilities using three example cases, covering the types of content and design feedback that can be an input to \name{} for generating infographics (Figure \ref{fig:example_cases};). The three input scenarios (indicated by rows) include (1) markdown content with no design input on the canvas, (2) markdown content with a pivot element placed on the canvas, and (3) markdown content with a pivot element and a hand drawing for the VIF layout.

For Example 1, \name{} is able to generate free designs varying the best-ranked VG designs, VIF layouts, and connection designs. 
For Example 2, the markdown content includes five VGs and a pivot graphic design input on the canvas. The recommendations, in this case, include a fixed pivot graphic, whereas the VIF layout, VG designs, and connection are ranked based on the content and the pivot element position. 
It shows that the tool can generate infographic designs based on the shape of the pivot element. 
%
Finally, for Example 3, there is an additional design input of a VIF layout hand drawing along with the same input from Example 2. In this case, \name{} generates infographics by ranking the VIF layout, VG designs, and connection designs adhering to a fixed pivot graphic and VIF layouts similar to the hand drawing with all other design components varying in each infographic. 
Based on the results from all the example inputs, we can conclude that our framework is able to generate infographic designs for specific design scenarios.


\subsection{Evaluating Design Experience with User Study}
\label{sec:user-study}

We further evaluated \name{} with real users for its ability to support the infographics design task. This study also aimed at investigating our framework's support for multiple factors of creativity. 
We did not conduct a comparative controlled study because we could not find comparable baselines that are publicly available to deploy and use. For commercial software, manual tools such as Adobe Illustrator is out of the scope of this work. Probably the most promising baseline is Microsoft PowerPoint, but it lacks the flexibility to control various design components (as indicated in Table \ref{tab:comp_hyp}), which is not a fair comparison. Thus, instead, this study's primary goal was to explore the strengths and weaknesses of \name{}, complementing the comparison described in Section \ref{ss:ach}.
 
\subsubsection{Study Setup}
\textbf{Participants.} 
We recruited 10 participants (five males and five females, aged between 25 and 35 years) via social media and mailing lists. The participants were carefully chosen to be designers with different expertise levels. Of all the participants, five have over three years of designing experience with various mainstream tools like Adobe Photoshop and Adobe Illustrator, categorized as \textit{experts} for this study; the other five have zero to less than a year's experience in designing infographics, categorized as \textit{non-experts} for this study. 
Out of the total of five experts, two were Ph.D. students working in Data Visualization, and three work in the design industry. All the five non-experts were graduate students chosen based on their design experience.

\textbf{Task and Procedure.} 
We initially familiarized the participants with the concepts of Infographics and related terminologies, such as the definition of VG, VIF, Pivot Graphic, and Connections for our framework. Next, we showed a few examples of infographics chosen from our dataset to familiarize them with infographic designs. The participants were then allowed to experiment with \name{} and ask clarifying questions regarding the tool. 
The task was to generate an infographic design from the content and design input of choice and evaluate the design support features of \name{}. We also provided sample markdown inputs to users, and they had an option to either generate infographics from the content of their choice or use the provided examples.  

After the participants were satisfied with their generated or filtered infographic designs, we conducted a short semi-structured interview to collect qualitative feedback. 
We asked the participants to rank their experience with \name{} on a five-point Likert scale. The questionnaire was based on a total of six factors: \textit{Enjoyment}, \textit{Exploration}, \textit{Expressiveness}, \textit{Results worth the effort}, \textit{Ease of use}, and \textit{Workflow}. Four of these factors were taken from the work by Cherry \textit{et al.} \cite{cherry2014quantifying} for quantifying the creativity support for design tools. Two specific factors, \textit{Effort} and \textit{Workflow}, were added in the questionnaire to specifically evaluate \name{} for its support in infographics design and exploration tasks. 
The whole study took about 45 minutes for each participant.

\subsubsection{Questionnaire Results}

\begin{figure*}[tb]
    \centering
        \includegraphics[width=\textwidth]{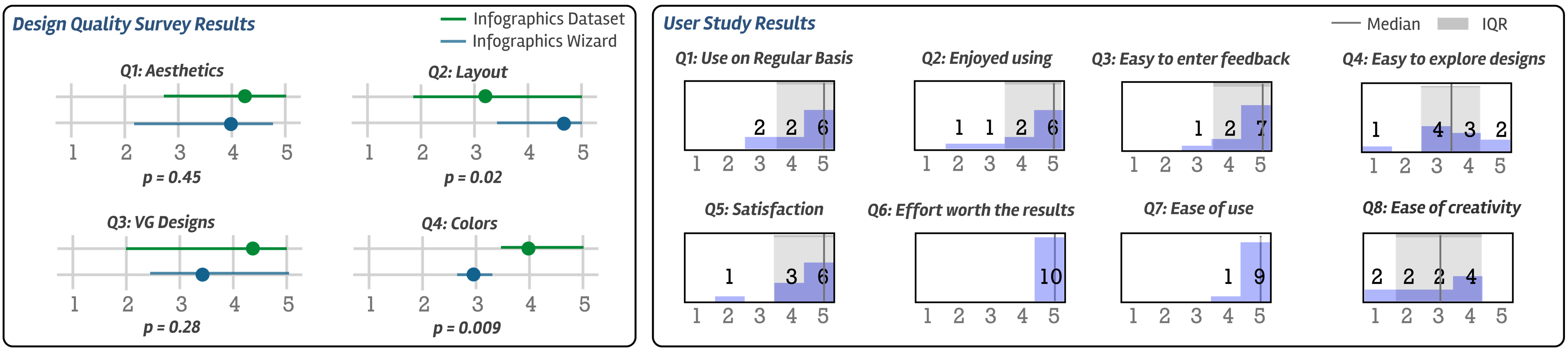}
        \vspace{-6mm}
    \caption{Evaluation results from the user survey (left) and the user study (right). Left is the design quality survey results showing the rating distribution (5-point scale) for the four questions to compare infographic design quality of \name{} with the baseline dataset. These distributions are compared using Wilcoxon test with p-values shown below each question. Right is the user study results showing rating distributions of the 5-point Likert scale, consisting of eight questions measuring various aspects of \name{}.} 
    \label{fig:case_study}
\end{figure*}

As shown in Figure \ref{fig:case_study}-right (User Study Results), on average, eight out of 10 participants rated \name{} with a score of 4 and 5 for every question. Overall, our tool received 100\% \textit{Agree votes} (scores of 4 and 5) on Q6 and Q7. Similarly, the second highest rated questions were Q1 and Q3. Besides the general results, we separately collected feedback for the low-scoring (scores of 1 and 2) features from particular participants. 
Q8 was the weakest scoring question from this study because four of the experts were trying to generate designs out of the scope of \name{}. Since our framework supports the designing of the majority (82\%) infographic designs derived from the formative study (Section \ref{ss:survey}), we interviewed the experts for their feedback and possible future directions of improvement in our work. More details about these exceptional cases are provided in the supplementary materials. 



\subsubsection{Interview Results with Experts}
To further validate our framework for its design capabilities from an expert's eyes, we conducted semi-structured interviews with two expert designers (\textit{E1} and \textit{E2}) from the user study to collect in-depth qualitative feedback about our framework. The interview also helped compare \name{} with existing design tools as both the participants had more than five years of experience designing infographics using the existing design tools like Adobe Photoshop and Figma. We discuss our results organized by the themes of the questionnaire questions in the following text.

\textbf{Ease of usage (Q7).} Both the participants found our framework easy to use for designing infographics. E1 explained \textit{``I really liked the simplistic design of the tool with a filtering and exploration panel on the side. It is effortless to navigate back and forth from design to exploration, similar to PowerPoint Design Ideas.''} E2 added that \textit{``I like the interface and the usage flow from left to right, moving from the content section on the left to the canvas and then finally the exploration section on the right.''}

\textbf{Usability (Q1, Q2, and Q6).} Both the experts found that the tool is handy in exploring different infographic designs. E1 commented that \textit{``I love the idea of separating the infographic content from design. The tool is really fast in generating infographic design recommendations for a given content.''} Similarly, E2 commented that \textit{``I can use \name{} as an exploration tool before designing very complex infographics to get new ideas. The fact that we can export final infographic designs as SVGs makes this tool very useful for designers.''}

\textbf{Quality of infographics (Q4, Q5, and Q8).} The participants were delighted with the variety of recommendations and the designs generated by our tool. E1 commented \textit{``The variety of VGs and Layouts is handy for exploration. I don't have to worry about finding the right template even for many VGs, which is a big problem in existing datasets. There are only a limited number of templates available for cases when the number of VGs exceeds 7.''} 
On similar lines, E2 commented about the designs generated for very complex layouts, saying \textit{``Now I can use the existing design of VGs to extend to very complex layouts. This was hard to achieve with existing infographic design tools as sometimes, we need an infographic design with a very unique layout. Most of the time, exploring infographic designs in such complex cases is impossible, and the designers have no ground truth to compare their work.''} E2 also commented about the creativity of the generated infographics saying, \textit{``I would have probably generated better infographics using Adobe Illustrator in some cases.''} However, they agreed that the amount of effort and time required to generate infographics would have increased. Instead, both participants suggested that they can export the designs and later edit them with Adobe Illustrator. 

\textbf{User feedback to the interface (Q3).} Both the participants were happy with the results generated after experimenting with our tool's design feedback feature. E1 commented about the pivot graphic functionality, saying, \textit{``I was surprised by the accuracy and speed of the final designs changing almost instantly as the pivot element is updated. It will be interesting to see the designs when this framework supports multiple pivot elements in the future.''} E2 commented, \textit{``I haven't seen any automatic infographic design tools where we can input so many design constraints. The ability to control the VG, VIF flow, connection designs, and even the color pallets gives the designers immense control over the design task. The tool is excellent in using the existing design pieces from large infographics datasets to generate very unique designs.''} 

\subsection {Evaluate Infographic Design Quality Survey Study}
We also performed a survey study to evaluate the design quality of infographics generated by \name{}. The primary goal of this study was to compare the design quality of infographics generated by our tool and the infographics collected during the formative study (Section \ref{ss:survey}). 


\textbf{Participants.}
We recruited 30 participants (20 males, 10 females, aged between 22 and 35 years) through social media and mailing lists for this study. The participants were chosen to be designers with different expertise levels similar to our in-lab user study. Out of the total 30 participants, 10 were experts (5 designers in the industry and 5 Ph.D. students working in data visualization), and 20 non-experts (less than a year's experience in designing). 

\textbf{Dataset.}
The dataset for the user study consisted of a total of 50 infographics, out of which 25 were generated by the participants using \name{} in the user study (Section \ref{sec:user-study}), and 25 images were chosen from the infographics dataset from the formative study (Section \ref{ss:survey}). To prevent any design biases in the images, the 25 images chosen from the formative study consisted of the same VIF distribution as the 25 images generated from \name{}. 

\textbf{Task and Procedure.}
After familiarizing with the terminologies of infographics, each participant was required to rate all 50 images on a 5 point scale (higher is better). The questionnaire consisted of four aspects: Aesthetics, Layout, VG Design, and Colors as shown in Figure \ref{fig:case_study} (left) - Design Quality Survey Results. Aesthetics covered the overall design quality of the infographic as perceived by the participants. The layout included the quality of VIF layouts of the VGs with respect to the overall design of the infographic. Similarly, VG design involved ratings of the quality of VGs, and finally, the color scheme was rated measuring the quality of overall color distribution in the infographic. The order of the infographics presented to the participants was randomized for each session.




\textbf{Results.} Figure \ref{fig:case_study}-left (Design Quality Survey Results) compares the scores received across the four aspects in both the original infographics from the formative study and the infographics generated from \name{} using the Wilcoxon test.
For aesthetics, Infographic Wizard's results and experts' designs were rated at par. This is encouraging because it indicates that our tool enabled many non-expert designers to achieve similar quality in infographic design on aesthetics. 
For layouts, designs by Infographic Wizard received a significantly higher rating, compared to those in the dataset. This verifies the utility of our tool and the goals of our framework in automating layout design based on VIF.

The third question is related to the VG designs, while the average rating on VG design is lower for \name{}, there is no significant difference between the two conditions, which indicates that Infographic Wizard can produce designs that have similar quality as hand-drawn infographic VGs. This is can be explained by the fact that \name{} uses VGs extracted from real-world datasets which are then ranked based on VIFs, thus producing close to hand-drawn results. Finally, for color schemes, original infographics have better color ratings than the generated infographics. This is because the users can only choose colors from pre-existing templates in \name{} as of now. We have decided to diversify the color schemes and add more flexibility in this aspect as a part of our future work.
Overall, the above results show that \name{} is useful and effective to assist expert and non-expert designers with generating the most common infographic designs. 


\section{Discussion, Conclusion, and Future Work}
\label{s:conclusion}
In this paper, we have presented an infographics authoring for structured and flow-based designs and exploration framework for a given user-input content. Our framework flexibly supports both fully automated infographic design with no design input and semi-automated infographic generation with design feedback, supporting designers with various experience levels. 

Overcoming the limitations of the existing design tools, our framework provides a comprehensive and sustainable infographic generation solution, realized in our tool, \name{}. Since we break down the infographic generation task into three independent pipeline steps, improving any one of these pipeline stages impacts the quality of the generated infographics directly. For example, we can only improve the VIF layout generation algorithm in the future without worrying about VG design and connection design rankings while still improving the generated infographic designs, or we can add a semantic similarity term in Equation~\ref{eq:layout_rank} to capture semantics of the infographics. Further, designers can also generate and explore infographics for very rare VG or VIF layout designs, which are not commonly found in the existing infographics templates. Our framework separates the content from the design aspect of infographic generation and reuses the design components from existing datasets to create new designs based on the user content. Hence, designers can now generate infographics for very complex layouts and even custom VG designs with minimal effort. We created an interface, \name{} to implement our framework, which can be used for design exploration. The infographic recommendations can be exported as SVG files for further low-level editing and fine-tuning with existing design tools like Adobe Illustrator. This feature is crucial in reducing the design time to generate very complex infographic designs.  

Besides the effectiveness of our framework, there remain limitations in some aspects regarding infographic design generation. We agree more complex IGs and some visualization types are not covered here. There is a class of infographics that do not follow the general norm of placing information inside similar VGs, which cannot be designed with our tool. Factors like scaling of VGs inside an infographic, using multiple VG designs, complex multi-object pivot elements, and complex connection designs are not yet supported in our framework. From the design aspect, \name{} currently only supports a single pivot element. Also, for the VIF layout hand drawing feature of the canvas, the highly ranked matching VIF layouts, in some cases, do not match the users drawing direction since we do not use the drawing direction as an input to rank closely related VIF flows. However, our framework is the first small step towards automating infographic design helping designers to create and explore most generic designs in no time. There is a trade-off between interface simplicity (IW) and advanced creativity support (Illustrator), we design IW for users who want to quickly generate reasonable IGs. 

Following up on the limitations, we have included these tasks as a part of our future work. We also plan to expand our VG and VIF datasets by adding more infographics from different sources since this will directly impact the design quality of the infographics generated by our framework. Another part of our future research is to generate more generic infographics not following the norm of a single VG and have multiple coherent VG designs. 
Moreover, we wish to conduct more user evaluations to investigate our approach's advantages and disadvantages thoroughly. We plan to deploy \name{} with real users in a longer-term study to collect in-depth feedback and usage scenarios.

\section{Acknowledgments}
We would like to thank the anonymous reviewers for their valuable comments. This work was made possible in part thanks to NSF awards IIS-1527200, CNS-1251137, CNS-1302246, CNS-1305360, CNS-1622832, CNS-1650499, CNS-1730726, and NSERC Discovery Grant.

\bibliographystyle{eg-alpha-doi}
\bibliography{references}
\end{document}